\documentclass[prd,twocolumn,amsmath,amssymb,floatfix, superscriptaddress]{revtex4}
\usepackage{graphicx, epsfig, bm,amsmath,natbib}

\usepackage{color}
\usepackage{hyperref}
\usepackage{ifthen}
\usepackage{xstring}
\usepackage{graphicx,epsfig}
\usepackage{subfigure}
\usepackage{amssymb}
\usepackage{afterpage}

\def\be{\begin{equation}}
\def\ee{\end{equation}}
\def\ba{\begin{eqnarray}}
\def\ea{\end{eqnarray}}
\def\nn{\nonumber}

\begin{document}

\title{Generalized Slow Roll for Large Power Spectrum Features}

\author{Cora Dvorkin}
\affiliation{Kavli Institute for Cosmological Physics, Enrico Fermi Institute,
        University of Chicago, Chicago, IL 60637}
\affiliation{Department of Physics, University of Chicago, Chicago, IL 60637}

\author{Wayne Hu}
\affiliation{Kavli Institute for Cosmological Physics, Enrico Fermi Institute,
        University of Chicago, Chicago, IL 60637}
\affiliation{Department of Astronomy \& Astrophysics, University of Chicago, Chicago, IL 60637}

\begin{abstract}
We develop a variant of the generalized slow roll approach for calculating the curvature  
power spectrum that is well-suited for order unity deviations in power caused by 
sharp features in the inflaton potential.    As an example, we show that predictions 
for a step function potential, which has been proposed to explain order unity glitches in the
CMB temperature power spectrum at multipoles $\ell = 20-40$,
are accurate at the
percent level.   Our analysis shows that to good approximation there is a single source function that is 
responsible for observable features and that this function is simply related to the
local slope and curvature of the inflaton potential.  These properties should make
the generalized slow roll approximation useful for inflation-model independent studies
of features, both large and small, in the observable power spectra. 
\end{abstract}
\maketitle

\section{Introduction}

The ordinary slow roll approximation provides a model-independent technique for
computing the initial curvature power spectrum for inflationary models where the
scalar field potential is sufficiently flat and slowly varying.  Such models lead
to curvature power spectra that are featureless and nearly scale invariant 
(e.g. \cite{Lidsey:1995np}).

On the other hand, features in the inflaton potential  produce features in the
power spectrum.   
Glitches in the observed temperature power spectrum of the cosmic microwave
background (CMB) \cite{Bennett:2003bz}
 have led to recent interest in exploring such models 
(e.g.~\cite{Peiris:2003ff,Covi:2006ci,Hamann:2007pa,Mortonson:2009qv,Pahud:2008ae,Joy:2008qd}).    To explain the glitches as other
than statistical flukes, these models
require order unity variations in the curvature power spectrum across about an $e$-fold
in wavenumber. 

Such cases are typically handled by numerically solving the
field equation on a case-by-case basis (e.g.~\cite{Adams:2001vc}).   For model-independent constraints and
model building purposes it is desirable to have a simple but accurate prescription that
relates features in the inflaton potential to features in the power spectrum (cf.~\cite{Hunt:2004vt,Habib:2004kc,Joy:2005ep,Kadota:2005hv}).

The generalized slow roll  (GSR) approximation was introduced by Stewart \cite{Stewart:2001cd} to overcome some of the problems of the ordinary slow roll approximation
for potentials with small but sharp features.  In this approximation, the ordinary slow
roll parameters are taken to be small but not necessarily constant.
  In this paper we examine and extend
the  GSR approach
for the case of large features where the slow-roll parameters
are also not necessarily small.  

In \S \ref{sec:GSR}, we review the GSR approximation and develop the variant
for large power spectrum features.   In the Appendix, we compare this variant to other GSR approximations
in the literature \cite{Stewart:2001cd,Choe:2004zg,Dodelson:2001sh,Kadota:2005hv,Gong:2005jr}.   We show that our variant provides both the most
accurate results and is the most simply related to the inflaton potential. 
In \S \ref{sec:applications}, we show how 
this technique can be used to develop alternate inflationary models to explain
a given observed feature.   We discuss these results in \S \ref{sec:discussion}.

\section{Generalized Slow Roll}
\label{sec:GSR}

The GSR formalism was developed to calculate the curvature power spectrum for
inflation models in which  the usual slow roll
parameters, defined in terms of time derivatives of the inflaton field $\phi$ and the
expansion rate $H$,
\begin{eqnarray}
\epsilon_H & \equiv & {1\over 2}\left( {\dot \phi \over H} \right)^2\,, \nonumber\\
 \eta_H& \equiv &-\left( \ddot \phi \over H\dot \phi \right)\,,
 \label{eqn:slowroll}
\end{eqnarray}
are small but $\eta_H (=-\delta_{1})$ is not necessarily constant.  In these models, the third slow-roll parameter
\begin{equation}
\delta_2={\dddot \phi \over H^2\dot{\phi}}\,,
\label{eqn:delta2}
\end{equation}
can be large for a small number of $e$-folds
\cite{Stewart:2001cd,Choe:2004zg,Dodelson:2001sh}.  Here 
and throughout we choose units where the reduced Planck mass $(8\pi G)^{-1/2}=1$. 

We study here the more extreme case where $\eta_H$ is also allowed to become large for a fraction of an $e$-fold.   These models
lead to order unity deviations in the curvature power spectrum.  As we shall
see, different implementations of the GSR approximation perform very differently for such 
models.

\begin{figure}[tbp]
\includegraphics[width=0.5\textwidth]{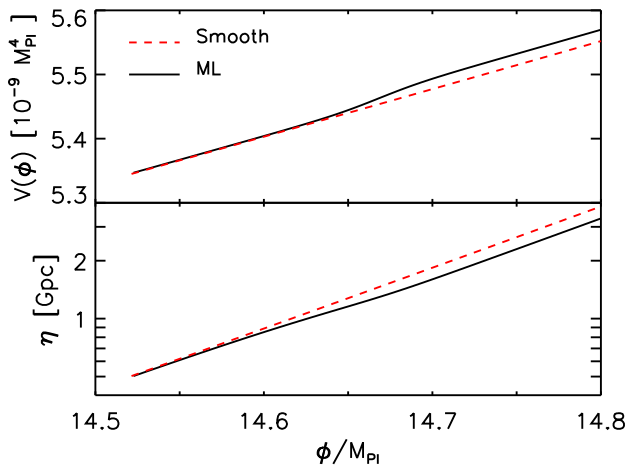}
\caption{Upper panel: inflationary potential with a step from Eq.~(\ref{eq:step_potential}) with
parameters that maximize the WMAP5 likelihood  (ML, black/solid) and an $m^2 \phi^2$ potential that matches the WMAP5 normalization (smooth, red/dashed).  Lower panel: conformal time to the end of inflation as a function of the value of the field. }
\label{plot:potential_time_bestfitmodel}
\end{figure}

An example
of such a case is a step in the 
inflaton potential of the form 
$V(\phi)=m^2_{\rm eff}(\phi)\phi^2/2$, where the effective mass of the inflaton potential is given by \cite{Adams:2001vc} 
\be\label{eq:step_potential}
m_{\rm eff}^2(\phi) = m^2 \left[1+c\tanh\left(\frac{\phi-b}{d}\right)\right]\,.
\ee
This form for the potential has been shown to be a good description of 
large features in the temperature power spectrum at $\ell \sim 20-40$ tentatively seen in
the WMAP data
 \cite{Covi:2006ci,Hamann:2007pa}.   The maximum likelihood (ML)
 parameters values for WMAP5 are $b=14.668$, $c=1.505\times10^{-3}$,
 $d = 0.02705$ and $m=7.126\times10^{-6}$ \cite{Mortonson:2009qv}.
 The potential for this choice of parameters is shown in Fig. \ref{plot:potential_time_bestfitmodel} (upper panel).
 For comparison we also show the best fit smooth model ($c=0$) with $m=7.12\times10^{-6}$.
 Since it will be convenient to express results in terms of physical scale instead of
 field value, we also show in the lower panel the relationship to 
 the conformal time to the end of inflation
 $
 \eta=\int^{t_{\rm end}}_t dt'/a.
 $
Note that $\eta$
is  defined to be positive during inflation.  The two models have comparable power at
wavenumbers $k \sim \eta^{-1} \sim 0.02$ Mpc$^{-1}$.

\begin{figure}[tbp] 
\includegraphics[width=0.5\textwidth]{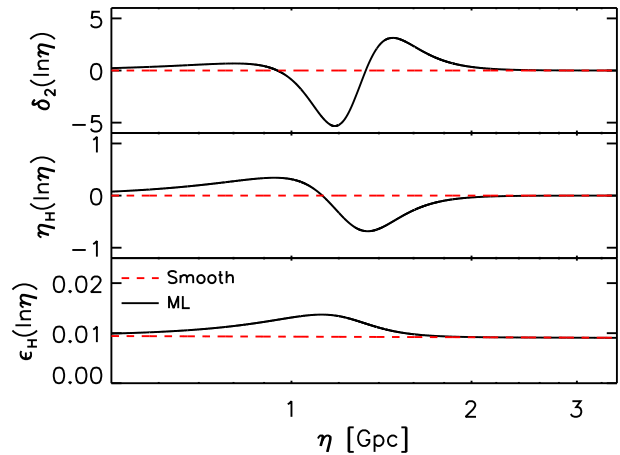} 
\caption{Slow-roll parameters $\epsilon_H$, $\eta_H$ and $\delta_2$ for the
two models of Fig.~\ref{plot:potential_time_bestfitmodel}: ML step model (black/solid) and smooth model (red/dashed).
}
\label{plot:slow_roll_parameters_bestfitmodel}
\end{figure}

 The slow-roll parameters for these models as a function of $\eta$ are shown in Fig. \ref{plot:slow_roll_parameters_bestfitmodel}.  Notice that $\epsilon_H$ remains
 small in the ML model though its value changes fractionally by order unity.  
 On the other hand, $\eta_H$ is of order unity and $\delta_2$ is greater than unity
 in amplitude
 in this model around $\eta \sim 1$ Gpc when the inflaton rolls across the feature.

\subsection{Exact Relations}

It is useful to begin by examining the exact equations and solutions. The exact equation of motion of each $k$-mode of the inflaton field is given by
\be
{d^2 u_k \over d\eta^2} +(k^2-{1 \over z} {d^2 z \over d\eta^2})u_k=0\,,
\ee
where
\begin{equation}
z= {f \over 2\pi \eta} \,, 
\quad
f  = 2\pi {{\dot\phi}a \eta \over H}\,.
\end{equation}

The field amplitude is related to  the curvature power spectrum  by
\be
\Delta^2_{\cal R}(k) = {k^3 \over 2\pi^2} \lim_{k\eta \rightarrow 0}\left| \frac{u_k}{z}\right|^2 \,.
\ee
Following  \cite{Stewart:2001cd},
we begin by transforming the field equation into  dimensionless
variables $y = \sqrt{2 k}u_k$, $x = k\eta$\be
{d^2 y \over d x^2} + \left( 1 - {2 \over x^2} \right) y = {g(\ln x) \over x^2} y \,,
\label{eqn:muktransformed}
\ee
where 
\be 
g = {f'' - 3 f' \over f}\,.
\ee
Primes here and throughout are derivatives with respect to $\ln \eta$. 

The functions $f$ and $g$ carry information about deviations from perfect slow
roll $\epsilon_H=0$, $\eta_H=0$ and $\delta_2=0$.  
Specifically, without assuming that these three parameters are small or  slowly varying
\begin{eqnarray}
\label{eqn:fslowroll}
f^2 &=& 8\pi^2 {\epsilon_H \over H^2} (a H\eta)^2 \,, \nonumber\\
{f'\over f} &=& -aH\eta(\epsilon_H -\eta_H) + (1-aH\eta) \nonumber\,, \\
{f''\over f}   &=& 3 {f' \over f} + 2[ (aH\eta)^2 - 1] \\
 \nonumber\\&&+(aH\eta)^2 [2\epsilon_H-3\eta_H  + 2\epsilon_H^2 - 4\eta_H \epsilon_H + \delta_2 ] \,,\nonumber
 \end{eqnarray}
and the dynamics of the slow-roll parameters themselves are given by
\begin{eqnarray}
{d \epsilon_H \over d\ln a} &=& 2\epsilon_H(\epsilon_H -\eta_H)\,,
\label{eqn:epsevol}\\
{d \eta_H \over d\ln a}  &=& \epsilon_H \eta_H + \eta_H^2 - \delta_2 \,.
\label{eqn:etaevol}
\end{eqnarray}
Moreover, these quantities are related to the inflaton potential by
\ba
({V_{,\phi}\over V})^2 &=& 2\epsilon_H{(1-\eta_H/3)^2 \over (1-\epsilon_H/3)^2}\,, \nonumber\\
{V_{,\phi\phi}\over V} &=& {\epsilon_H+\eta_H-\delta_2/3 \over 1-\epsilon_H/3}\,,
\ea
which in the limit of small and nearly constant $\eta_H$, $\epsilon_H$ return the ordinary
slow roll relations.

In general, there is no way to directly express the source function $g$  in terms of the potential
without approximation.  Here
we want to consider a situation where the feature in the potential is not
large enough to interrupt inflation and hence $\epsilon_H \ll 1$, but is sufficiently large
to make $\eta_H$ of order unity for less than an $e$-fold.  By virtue of
Eq.~(\ref{eqn:etaevol}), $|\delta_2| \gg 1$ during this time.   This differs from other
treatments which assume $|\eta_H| \ll 1$ and by virtue of 
Eq.~(\ref{eqn:epsevol}) a nearly constant $\epsilon_H$ \cite{Stewart:2001cd}. 

Even under these generalized assumptions there are some terms in $\eta_H$ and
$\delta_2$ that can be neglected.  For example,
even if $\eta_H$ is not small, it suffices to take
\begin{equation}
aH \eta - 1 = \epsilon_H + \epsilon_H {\cal O}(\eta_H) \,.
\end{equation}
This expression preserves the ordinary slow roll relations when $|\eta_H| \ll 1$.
When $\eta_H$ is not small, this quantity remains of order $\epsilon_H$ and
so is negligible compared with bare $\eta_H$ and $\delta_2$ terms.
Hence this approximation suffices everywhere.

\begin{figure}[tbp]
\includegraphics[width=0.45\textwidth]{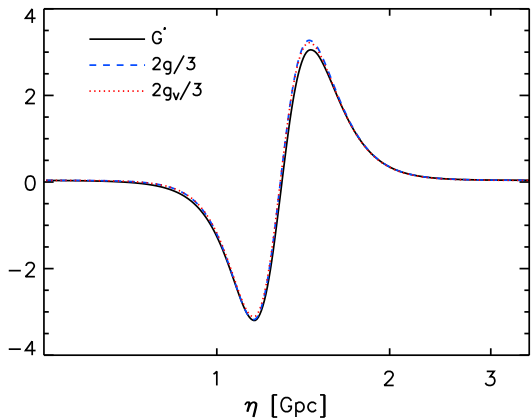}
 \caption{Source functions for the deviations from slow roll used in the
 GSR approximations: $2g/3$, $2 g_V/3$ and
  $G'$ (see \S \ref{sec:GSRL})  for the maximum likelihood model.   To
  good approximation $g=g_V$ which directly relates the source function to features
  in the inflaton potential. Likewise $G' \approx 2g_V/3$ and is most simply related to the
  curvature power spectrum for large deviations.    }
\label{plot:Gprime_g_gV}
\end{figure}

Following this logic,
we obtain
\be\label{eq:Gprime_potential_eta}
g=g_V +
\epsilon_H {\cal O}(\epsilon_H,\eta_H,\delta_2)  \,,
\ee
where $g_{V}$ is directly related to the potential
\begin{eqnarray}
g_{V} &\equiv& {9 \over 2}  ({V_{,\phi}\over V})^2 - 3{V_{,\phi\phi}\over V} 
\nonumber\\
& = &
 6\epsilon_H - 3\eta_H + \delta_2 
+
\epsilon_H {\cal O}(\epsilon_H,\eta_H,\delta_2) 
\,.
\label{eqn:gtoV}
\end{eqnarray}
As shown in  Fig. \ref{plot:Gprime_g_gV}, this relationship between the source function
 $g$  and features in the potential $V$ holds even for the
ML step potential.   Thus, if we can express the functional relationship between
 $g$ and the curvature power spectrum that is valid for large $g$ we can use
 features in the power spectrum to directly constrain features in the inflaton potential.

To determine this relation note that in the $x$ and $y$ variables
 the curvature is
$
{\cal R} = { x y / f} ,
$
and its power spectrum is
$
\Delta^2_{\cal R}(k)  = \lim_{x \ll 1} \left| {\cal R} \right|^2.
$
The LHS of Eq.~(\ref{eqn:muktransformed}) is simply the equation for scale invariant perfect slow roll and
is solved by
\begin{equation}
y_0(x)  = \left( 1 + {i \over x} \right) e^{ix}\,,
\end{equation}
and its complex conjugate $y_0^*(x)$.    
An exact, albeit formal solution to the field equation can be constructed with the
Green function technique \cite{Stewart:2001cd}
\begin{equation}
y(x) = y_0(x) - \int_x^\infty {du \over u^2} g(\ln u) y(u) {\rm Im}[ y_0^*(u) y_0(x)  ]\,.
\label{eqn:formalfield}
\end{equation}
The solution is only formal since $y$ appears on both the left and right hand side
of the equation.    The corresponding formal solution for the curvature power spectrum can be
made more explicit by parameterizing the source $y(u)$ as
\begin{equation}
y(u) = F_{\rm R}(u) {\rm Re}[y_0(u)] + i F_{\rm I}(u) {\rm Im}[y_0(u)]
\end{equation}
so that
 \begin{eqnarray}
 \label{eqn:gsrfieldlimit1}
\lim_{x \ll 1} {  (x y) } &= & i - {i \over 3}  \int_x^\infty {du \over u} { x^3 \over  u^{3}} F_{\rm I}(u) g(\ln u)
\\&& + 
{i \over 3} \int_x^\infty {d u \over u} W(u) F_{\rm I}(u) g(\ln u) 
\nonumber\\&& +
{1 \over 3} \int_x^\infty {d u \over u } X(u) F_{\rm R}(u) g(\ln u)
\nonumber\\&& +
{x^3 \over 9} \int_x^\infty {d u \over u} W(u) F_{\rm R}(u) g(\ln u) + {\cal O}(x^2)\,,\nonumber
\end{eqnarray}
where 
\begin{eqnarray}
W(u) &\equiv& -{3 \over u}{\rm Im}[y_0(u)] {\rm Re}[y_0(u)] \nonumber\\
          & =& {3 \sin(2 u) \over 2 u^3} - {3 \cos 2 u \over u^2} - {3 \sin(2 u)\over 2 u} \,,
        \nonumber\\
X(u) &\equiv& { 3 \over u}{\rm Re}[y_0(u)] {\rm Re}[y_0(u)]  
\nonumber\\
&=& -{3 \cos(2u) \over 2 u^{3}} - {3 \sin(2 u) \over u^{2}} 
\nonumber\\&& + {3 \cos(2 u) \over 2 u} +
{3 \over 2 u^{3}}(1+ u^{2}) \,.
\end{eqnarray}
Note that $\lim_{u\rightarrow 0} W(u)=1$ and $\lim_{u \rightarrow 0} X(u) = u^3/3$ and 
we have utilized the fact that
\begin{equation}
{\rm Im}[y_0(u)]{\rm Im}[y_0(u)] = {1 + {1 \over u^2} } - {u\over 3} X(u) 
\end{equation}
goes to $1/u^2$ in the limit $u \rightarrow 0$.

Finally, the curvature power spectrum becomes
\be
\Delta^2_{\cal R}(k)  = \lim_{x \ll 1}x^2 {[{\rm Im}(y)]^2 + [{\rm Re}(y)]^2 \over f^2}  \,,
\label{eqn:curvatureintegral}
\ee
with  $y$ given by Eq.~(\ref{eqn:gsrfieldlimit1}).

\subsection{GSR for Small Deviations}
\label{sec:GSRS}

The fundamental assumption in GSR is that one recovers a good solution by setting
$F_{\rm I}(u)=F_{\rm R}(u)=1$ in the formal solution for the field fluctuations in
Eq.~(\ref{eqn:gsrfieldlimit1}).  Equivalently, $y(u) \rightarrow y_0(u)$ in the source term on 
the RHS of  Eq.~(\ref{eqn:muktransformed}).   Note that this does {\it not} necessarily require that
 $g$ itself is everywhere much less than unity.  For example, modes that encounter a
 strong variation in $g$ while deep inside the horizon do not retain any imprint of the
 variation and hence the GSR approximation correctly describes the curvature they
 induce.

 \begin{figure}[tbp]
\includegraphics[width=0.45\textwidth]{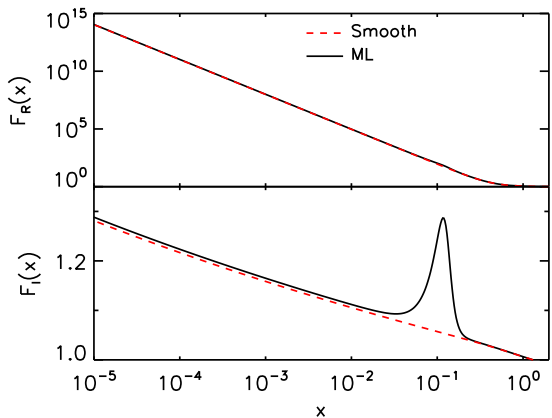}
 \caption{Ratio of field solution $y$ to the scale invariant approximation $y_{0}$.
 Upper panel: real part $F_{\rm R}$  for a smooth case (red/dashed line), and for the maximum likelihood model (black/solid line), both at $k=10^{-4}$ Mpc$^{-1}$. Lower panel: imaginary part $F_{\rm I}$  for the same models.}
\label{plot:FR_I}
\end{figure}

In Fig.~\ref{plot:FR_I}, we show an example of $F_{\rm I}$ and $F_{\rm R}$ for
a mode with $k=10^{-4}$ Mpc$^{-1}$ for both the ML and smooth models.   For the ML model,
this mode is larger than the horizon when the inflaton 
crosses the feature.  
Note that even in the smooth model, the two functions deviate substantially from unity 
at $x \ll 1$.  In fact, they continue to increase indefinitely 
after horizon crossing and $F_{\rm R} \propto x^{-3}$ diverges to compensate
for $|{\rm Re}(y_0)| \propto x^2$.
For the ML model, even $F_{\rm I}$ deviates strongly from unity during the crossing of 
the feature at $x \sim 0.1$.

 The impact that these deviations have on the curvature spectrum can be better understood
 by reexpressing the various contributions in a more compact form.
First note that
\begin{eqnarray}
\lim_{x \ll 1} {x^3 \over 3} \int_x^\infty {du \over u} u^{-3} g(\ln u)
&=& -{1 \over 3} \left( { f' \over f} \right) \,,
 \end{eqnarray}
 and so Eq.~(\ref{eqn:gsrfieldlimit1}) becomes 
 \begin{equation}
 \lim_{x \ll 1} { | {\cal R}_{\rm GSRS} |}  ={1 \over f} \left[  1 + {1 \over 3} {f' \over f} + 
{1 \over 3} \int_x^\infty {d u \over u} W(u)g(\ln u) \right] \,,
\label{eqn:GSRbasic}
\end{equation}
where note that we have dropped the Re($xy$) contribution
since it adds in quadrature to the power spectrum and hence is second order in $g$.   We call this the ``GSRS" approximation
for the curvature power spectrum $\Delta_{\cal R}^2 = 
\lim_{x \ll 1} | {\cal R}_{\rm GSRS}|^2$
 given its validity for small
fluctuations in the field solution from $y \rightarrow y_0$. 

The choice of $x$ is somewhat problematic \cite{Stewart:2001cd}.   From Fig.~\ref{plot:FR_I},
we see that taking $x$ too small will cause spurious effects since
$F_{\rm I}$ increases as $x$ decreases.  On the other hand, $x$ cannot be chosen to
be too large for the ML model since it will cause some $k$ modes to have their curvature calculated 
when the inflaton is crossing the feature.   Moreover, if $x$ is set to be some
fixed conformal time during inflation $\eta_{\rm min}$, then it will vary with $k$.  

\begin{figure}[tbp] 
\includegraphics[width=0.45\textwidth]{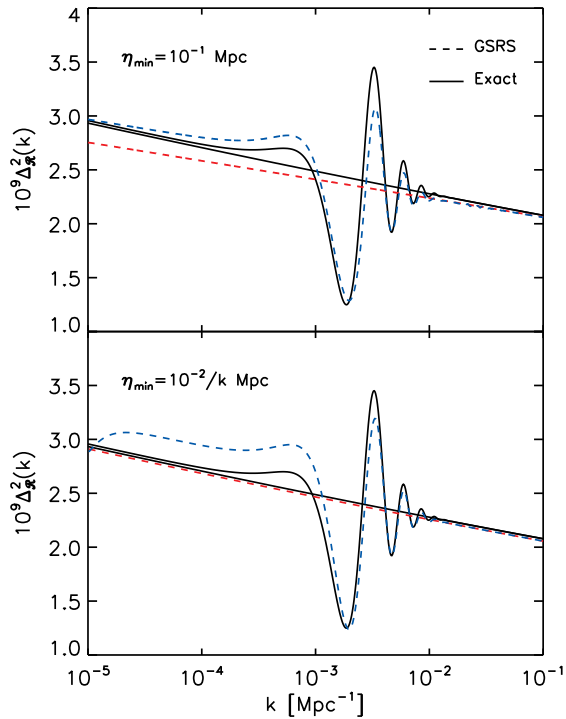}
 \caption{GSRS approximation to the curvature power spectrum (dashed lines) compared to the exact solution (solid lines) for a choice of $\eta_{\rm min}=10^{-1}$ Mpc (upper panel) and $\eta_{\rm min}=10^{-2}/k$ Mpc (lower panel).  The ML model is shown in blue and the smooth model in red for GSRS. }
\label{plot:GSRS_Muk_bestfitmodel}
\end{figure}

We illustrate these problems in Fig.~\ref{plot:GSRS_Muk_bestfitmodel}.  For
$\eta_{\rm min}=10^{-1}$ Mpc  (upper panel), GSRS underpredicts power at low $k$
for the smooth model and overpredicts it for the ML model. 
Agreement for the smooth model is improved by choosing $x = 10^{-2}$, i.e.~nearer to horizon crossing (cf.~Appendix for variants that take $x \approx 1$).   On the other hand, the agreement for the ML model becomes worse and has a spurious feature at $k \sim 10^{-5}$
Mpc$^{-1}$ where the inflaton is crossing the feature at $x=10^{-2}$.
In the next section, we shall examine the origin of the deviations from the exact solution 
and how a variant of the GSR approximation can fix most of them.

\subsection{GSR for Large Deviations}
\label{sec:GSRL}

When considering large deviations from scale invariance,
either due to sharp features in the potential or due to extending the calculation for
many $e$-folds after horizon crossing,  the first qualitative problem with the GSRS approximation of Eq.~(\ref{eqn:GSRbasic}) is that it 
represents a linearized expansion for a correction that is not necessarily small.   When
the correction becomes large, ${\cal R}_{\rm GSRS}$ can pass through zero leaving nodes
in the spectrum.  While this is not strictly
a problem for the ML model, it is better to have a more robust implementation of GSR
for likelihood searches over the parameter space. 

We can finesse this problem by replacing the linearized expansion 
$1+x$ by $e^x$ and write the power spectrum in the form 
\be
\ln\Delta_{\cal R}^2(k) = G(\ln\eta_{\rm min}) + {2 \over 3}\int_{\eta_{\rm min}}^{\infty}{d\eta\over\eta}W(k\eta)g(\ln\eta)\,,
\label{eqn:GSR1}
\ee
where
\be
G(\ln\eta)=\ln\left({1\over f^2}\right) + {2\over 3}{f'\over f}\,.\\
\label{eqn:Gdef}
\ee
This procedure returns the correct result at first order since $g$ and $f'/f$ are both
first order  in the slow-roll parameters (see Eq.~(\ref{eqn:fslowroll})).
We shall see below that it can be further modified to match the fully non-linear
result for superhorizon modes.

\begin{figure}[tbp]
\includegraphics[width=0.45\textwidth]{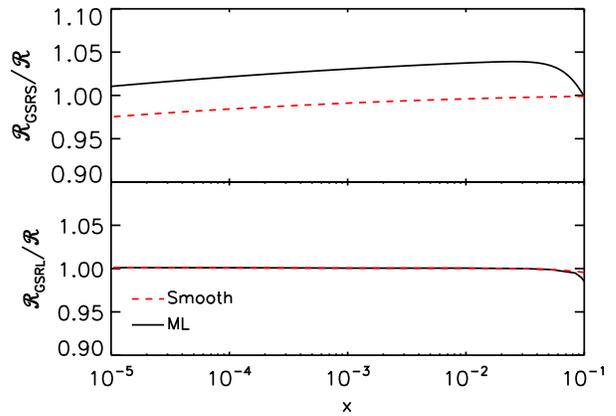}
 \caption{Curvature evolution after horizon crossing in the GSRS (upper panel) and GSRL (lower panel) approximations, both normalized to the exact solution. The ML model (black/solid line) and smooth model (red/dashed line) are both shown at $k=10^{-4}$ Mpc$^{-1}$.}
\label{plot:GSRS_GSRL}
\end{figure}

The more fundamental problem with GSRS is the deviation of the true solution $y$ from the scale
invariant solution $y_0$ when the mode is outside the horizon (see Fig.~\ref{plot:FR_I}).
The origin of this problem is that the exact solution requires the curvature ${\cal R} = xy/f$ to
be constant outside the horizon, independently of how strongly $f$ evolves.   Thus, if $f$ is allowed to vary significantly, either due to 
the large number of $e$-folds that have intervened since horizon crossing or due to 
a feature in the potential, then $y$ must follow suit and deviate from $y_0$ breaking the
GSRS approximation.

Fig.~\ref{plot:GSRS_GSRL} (upper panel) illustrates this problem.  Even for the smooth
model, the curvature is increasingly underestimated as $x \rightarrow 0$ .  With
the ML model, the crossing of the feature induces an error of the opposite sign.   For
 $x \sim 10^{-5}$ these problems fortuitously cancel but not for any 
fundamental or model independent reason.  

Given this problem, GSRS actually works better than one might naively expect.
For example at $k=10^{-4}$ Mpc$^{-1}$, 
even though $F_{\rm I} \sim 1.28$ at $x=10^{-5}$, the GSRS approximation gives a $\sim 2.5\%$ difference in the curvature and a $\sim 5\%$ difference
in the power spectrum with the exact solution for the smooth model instead of the 
$28\%$ and the $64\%$ differences one might guess. 
The main contribution to the GSRS correction from scale invariance is given by the integral
term in  Eq. (\ref{eqn:GSRbasic}), which is $\sim 0.25$ for the smooth case. Given that $F_I$ is a linear function in $\ln \eta$ and $g$ is slowly varying, 
we can approximate enhancement due to $F_I$ of the integral term 
by its average interval ($\sim 1.14$).  With this rough estimate we obtain an approximately $(1+0.25)^2/(1+0.25\times1.14)^2 \sim 5-6\%$ error in power  in 
agreement with the power spectrum result in Fig. \ref{plot:GSRS_Muk_bestfitmodel}.  

Furthermore although $F_{\rm R}$ diverges as $x^{-3}$ in Fig.~\ref{plot:FR_I}, the contribution
to the power spectrum of the real part of $y$ remains small. Its absence in the GSRS approximation produces
a negligible effect for modes that are larger than the horizon when the inflaton crosses
the feature.   The integrands for the real contribution contain either the function $X$, which
peaks at horizon crossing $x \sim 1$, or $x^3 W(u)$ which is likewise suppressed
at $x \ll 1$.   The correction adds in quadrature to the imaginary part and so it is intrinsically
a second order correction (see \S \ref{sec:iterative}).   
 For $k=10^{-4}$ Mpc$^{-1}$ its contribution to the power spectrum is $0.08\%$ of the
 power spectrum in the ML model.  
The fact that integrals over the deviation of $y$ from $y_0$ can remain small even
when neither $g$ nor the maximum of $y-y_0$ is small is crucial to explaining why the GSR approximation
works so well and why we can extend GSRS with small, controlled corrections.

Nonetheless these problems with GSRS are significant and exacerbated by the presence
of sharp features in the potential. 
The fundamental problem with GSRS is that its results depend on an arbitrarily chosen
value of $x \ll 1$, i.e.~${\cal R}$ is not strictly constant in this regime.   Phrased in terms of Eq.~(\ref{eqn:Gdef}) the problem is
 that $g$ is not directly related to $G$
but rather
\begin{equation}
{2 \over 3} g = G' + {2 \over 3}\left( {f' \over f}\right)^2 \,,
\end{equation}
where
\begin{equation}
G' = {d G \over d\ln \eta} = {2\over 3}({f''\over f} - 3{f'\over f} - {f'^2\over f^2}) \,.
\end{equation}
In GSRS, replacing $g$ with $3G'/2$ amounts to a second order change in the source function.   In fact even for the ML step function this change
is a small fractional change of the source everywhere in $\ln \eta$: it is small as the inflaton rolls past the feature since $|f''/f |\gg (f'/f)^2$ and
it is small before and after this time since $|f'/f| \ll 1$. 
In terms of the slow-roll parameters, this replacement is a good approximation if $\eta_H^2 \sim O(1)$ only where $|\delta_2| \gg 1$ and $g \approx \delta_2$ (see Eqs.~(\ref{eq:Gprime_potential_eta}) and (\ref{eqn:gtoV})).
\begin{equation}
G' = {2 \over 3} g + {2 \over 3} \eta_H^2 + \epsilon_H {\cal O}(\epsilon_H,\eta_H,\delta_2) 
\end{equation}
Moreover $G' \approx 2 g_V/3$ and remains
directly relatable to the inflaton potential through Eq.~(\ref{eqn:gtoV}).  For comparison
we show all three versions of the GSR source function in Fig.~\ref{plot:Gprime_g_gV}.

Nonetheless, 
the replacement
can have a substantial effect on the curvature once the source is  integrated over $\ln \eta$ because the difference is a positive
definite term in the integral.   Moreover, this cumulative effect is exactly what is needed to recover the
required superhorizon behavior.  Replacing $2g/3 \rightarrow G'$ in the power spectrum expression, we obtain \cite{Kadota:2005hv}
\be
\ln\Delta_{\cal R}^2(k) = G(\ln\eta_{\rm min}) + \int_{\eta_{\rm min}}^{\infty}{d\eta\over\eta}W(k\eta)G'(\ln\eta) \,,
\label{eqn:GSR2}
\ee
which we call the GSRL approximation.  The field solution corresponding to this
approximation, valid for $x\ll 1$, is given by
\begin{equation}
\lim_{x \ll 1} { | x y |} = \exp\left[{1\over 3} {f'\over f} + 
{1 \over 2} \int_x^\infty {d u \over u} W(u) G'(\ln u) \right]\,.
\label{eqn:gsrfieldlimit2}
\end{equation}

Now any variation in $f$ while the mode is  outside the horizon and $W(k\eta) \approx 1$
integrates away and gives the same result as if $\ln \eta_{\rm min}$ were set to be 
right after horizon crossing for the mode in question.  This can be seen more clearly
by integrating Eq.~(\ref{eqn:GSR2}) by parts \cite{Kadota:2005hv}
\be
\ln\Delta_{\cal R}^2(k)  = -\int_{\eta_{\rm min}}^{\infty}
{d\eta\over\eta}W'(k\eta) G(\ln\eta)\,.
\label{eqn:GSR3}
\ee
Since $- \int_0^\infty d\ln x W'(x) = 1$ and $\lim_{x\rightarrow 0} W'(x) \rightarrow 0$, 
the curvature spectrum does not depend on the evolution of $f$ outside the horizon.  
Moreover,  the integral
gets its contribution near $x\sim 1$ so for smooth functions $G(\ln\eta)$ we recover the slow roll expectation that 
\begin{equation}
 \ln\Delta_{\cal R}^2(k) \approx G(\ln \eta)\Big|_{k\eta \approx 1} \,.
 \label{eqn:OSR}
 \end{equation}
 If the slow-roll parameters are all small then 
the leading order term in Eq.~(\ref{eqn:OSR}) returns
the familiar expression for the curvature spectrum $\Delta_{\cal R}^2 \approx f^{-2} \approx H^2/8\pi^2\epsilon_H$ at $k\eta \approx 1$.
Choe et al.~\cite{Choe:2004zg} showed that Eq.~(\ref{eqn:GSR3}) is correct up to
second order in $g$ for $k\eta \ll 1$.  Here we show that it is correct for
arbitrary variations in $f$ and $g$ outside the horizon.
 
\begin{figure}[tbp]
\includegraphics[width=0.45\textwidth]{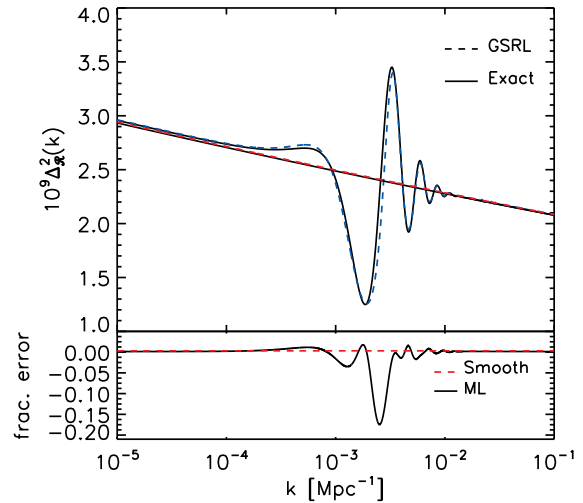}
 \caption{GSRL approximation to the curvature power spectrum.  Upper panel: approximation compared with the exact solution (solid lines) for the maximum likelihood model. Lower panel: fractional error between  the approximation 
 and the exact solution. }
\label{plot:GSRL_Muk_c0_bestfitmodel}
\end{figure}

 The superhorizon curvature evolution for $k=10^{-4}$ Mpc$^{-1}$ 
  corresponding to the GSRL approximation is
 shown in Fig. \ref{plot:GSRS_GSRL} (lower panel).  In the $x \ll 1$ domain of applicability
 of Eq.~(\ref{eqn:gsrfieldlimit2}), the curvature is now appropriately constant for
 both the ML and smooth models.   
 The net result is that the curvature power spectrum shown in Fig.~\ref{plot:GSRL_Muk_c0_bestfitmodel} is now a good match to the exact solution for
 low $k$.

\subsection{Power Spectrum Features}
\label{sec:slowrollparam}

We now turn to issues related to the response of the field and curvature 
for $k$ modes that are inside the horizon when the inflaton rolls across the feature. 
Fig.~\ref{plot:GSRL_Muk_c0_bestfitmodel} shows that the GSRL approximation 
works remarkably well for the ML model despite the fact that the power spectrum
changes by order unity there.  
The main problem is a $\sim 10-20\%$ deficit of power for a small range in $k$
near the sharp rise between the
trough and the peak.   

\begin{figure}[tbp]
\includegraphics[width=0.45\textwidth]{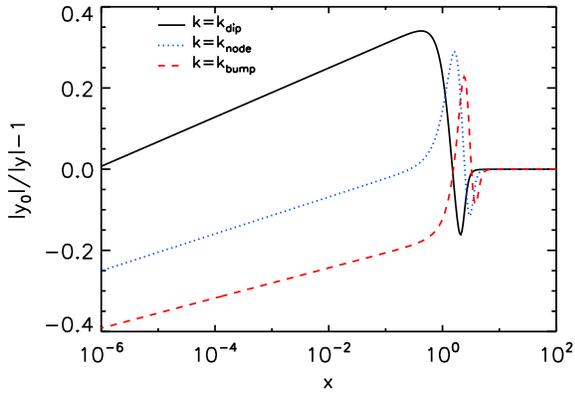}
 \caption{Fractional difference between $|y_{0}|$ and $|y|$ for the ML model at $k$ values at the dip, node and bump of the feature in the power spectrum (see text).}
\label{plot:y_over_y0_kdip_node_bump}  
\end{figure}

In Fig. \ref{plot:y_over_y0_kdip_node_bump}, we show the deviation of
the exact solution $y$ from the scale invariant $y_0$ that is at the heart of the GSR 
approximation.  The three modes shown, $k_{\rm dip}=1.8\times10^{-3}$ Mpc$^{-1}$, $k_{\rm node}=2.5\times10^{-3}$ Mpc$^{-1}$, $k_{\rm bump}=3.2\times10^{-3}$ Mpc$^{-1}$, correspond to the first dip, node and bump in the power spectrum of the ML model.

The first thing to note is that for higher $k$, the inflaton crosses the feature
at increasing $x$ where the deviations of $y$ from $y_0$ actually decrease.   Hence
the fundamental validity of the GSR approximation actually improves for subhorizon modes.
Combined with the GSRL approximation that enforces the correct result at $x \ll 1$, this
makes the approximation well behaved nearly everywhere.  

The small deviations from GSRL appear for modes that cross the horizon right around the
time that the inflaton crosses the feature.  It is important to note that the step potential
actually provides two temporal features in $g$ or $G'$ displayed in Fig.~\ref{plot:Gprime_g_gV}.  Each mode first crosses a positive feature
at high $\eta$ and $x$ and then goes through a nearly equal and opposite negative feature.    The end result
for the field amplitude or curvature is an interference pattern of contributions from both
temporal features.  For example, the peak in power is due to the constructive interference
between a  positive response to the positive feature and a negative response to the 
negative feature.   This suggests that one problem with the GSRL approach is
that it does not account for the deviation of the field $y$ from $y_0$ that accumulates through passing the
positive temporal feature when considering how the field goes through the negative feature.   This
is intrinsically a non-linear effect.

\begin{figure}[tbp]
\includegraphics[width=0.45\textwidth]{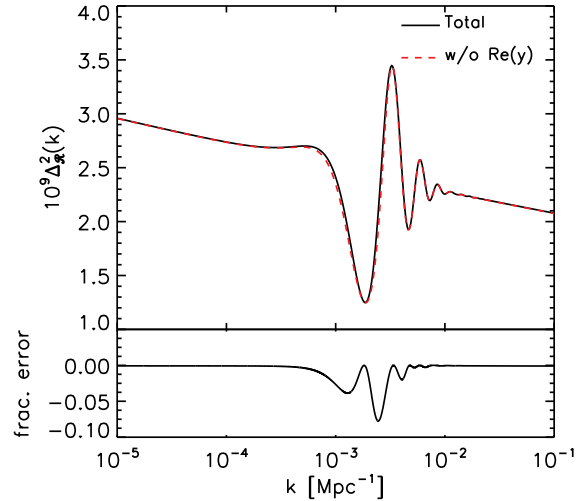}
\caption{Contribution of the real part of the $y$ field 
to the curvature power spectrum.  Upper panel: 
spectrum with and without the real part. Lower panel: fractional error between the two solutions.  }
\label{plot:Powersp_real_ML}
\end{figure}

 The final thing to note is that, since $g$ and $G'$ are of order unity as these modes
 exit the horizon, the real part of the field solution is not negligible.  Moreover,
 it contributes a positive definite piece to the power spectrum.   In Fig.~\ref{plot:Powersp_real_ML}, we show the result of dropping the real part from the
 exact solution.   Note that the fractional error induced by dropping the real part looks much
 like the GSRL error in Fig.~\ref{plot:GSRL_Muk_c0_bestfitmodel} but with $\sim 1/2$ the amplitude.

\subsection{Iterative GSR Correction}
\label{sec:iterative}

The good agreement between GSRL and the exact solution even in the presence
of large deviations in the curvature spectrum suggests that a small higher order correction may further improve the accuracy.  Moreover, the analysis in the previous section implies that there are two
sources of error: the omission of the field response from inside the horizon $x>1$ when
computing the response of the field to features at horizon crossing $x \sim 1$ and the
dropping of the real part of the field solution.

  Both of these contributions come in at
second order in the GSR approximation.
All first order GSR variants involve the replacement of  the true field solution $y$ with the scale invariant solution $y_0$
in Eq.~(\ref{eqn:muktransformed}).     This replacement can be iterated with successively
better approximations to $y$. 
  We begin with the GSRS approximation of replacing $y \rightarrow y_0$
to obtain the first order solution $y_1$.  We then replace $y \rightarrow y_1$ in the source
to obtain a second order solution $y_2$, etc.   

\begin{figure}[tbp]
\includegraphics[width=0.45\textwidth]{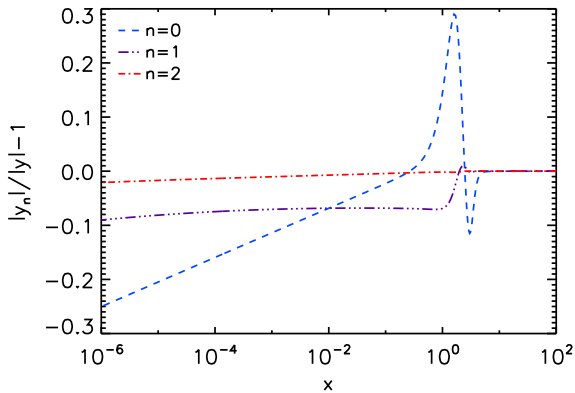}
\caption{Fractional difference between the exact ($y$) and $n$th order iterative solutions
($y_{n}$) for the ML model at $k=k_{\rm node}$ where the errors in the GSRL approximation are maximized. }
\label{plot:ratio_of_ys_iteration_atknode}
\end{figure}

\begin{figure}[tbp]
\includegraphics[width=0.45\textwidth]{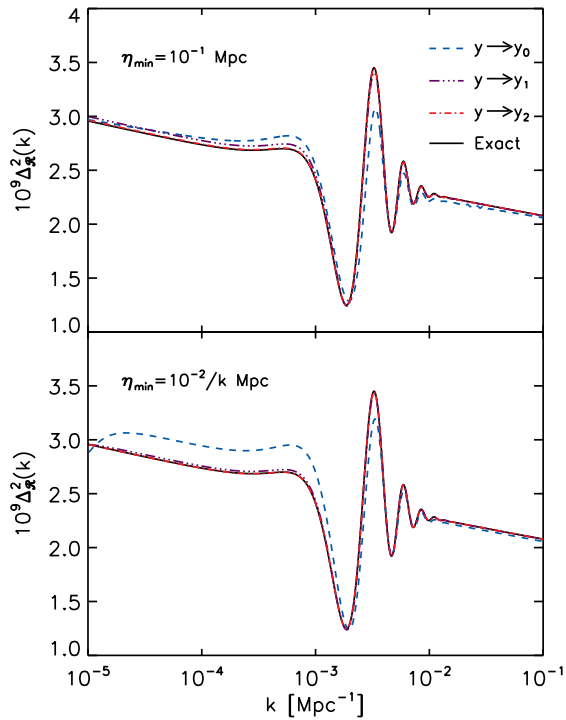}
\caption{Curvature power spectrum in the GSRS approximation for $\eta_{\rm min}=10^{-1}$ Mpc (upper panel) and $\eta_{\rm min}=10^{-2}/k$ Mpc (lower panel) when $y \rightarrow y_n$ in the GSRS source compared to the exact solution.}
\label{plot:GSRS_2iterations}
\end{figure}

We show the fractional error between the iterative solutions and the exact solution for $k=k_{\rm node}$ in Fig. \ref{plot:ratio_of_ys_iteration_atknode}, where the error in GSRL is roughly maximized.
As in the first order GSRS approach, the accuracy depends on the arbitrary choice of
$x = k \eta_{\rm min}$ when the curvature is computed.   The number of iterations required
for a given accuracy increases with decreasing $x$.  
We show the curvature  spectrum  in Fig. \ref{plot:GSRS_2iterations} for the same two choice of $\eta_{\rm min}=10^{-1}$ Mpc (upper panel) and $\eta_{\rm min}=10^{-2}/k$ Mpc (lower panel) as in Fig.~\ref{plot:GSRS_Muk_bestfitmodel}.  Note that in both cases, the
result has converged at the $0.5\%$ percent level or better to the exact solution within
three iterations.

Unfortunately the iterative GSRS approach is not of practical use in that each iteration requires essentially the same effort as a single solution of the exact approach.
On the other hand, 
rapid convergence in the iterative GSRS approach suggests that a nonlinear
correction to GSRL based on a second order expansion might suffice.   A second order GSRL approach differs conceptually from
the iterative GSRS approach in that it is formally an expansion in $g$ where in our case
$|g|\ll 1$ is not satisfied.   The iterative GSRS approach is exact in $g$ but expands in
$y-y_0$.   What makes a second order GSRL approach feasible is that the critical elements
involve time integrals over $g$ which can be small even if $g$ is not everywhere small.

Our strategy for devising a non-linear correction to GSRL is to choose a form
that reproduces GSRL at first order in $g$, is exact at second order in $g$, is simple to relate to the inflaton potential, and finally
is well controlled at large values of $g$.   
The second order in $g$ expressions for 
the curvature
are explicitly given in \cite{Choe:2004zg} and come about by both iterating the integral
solution in Eq.~(\ref{eqn:gsrfieldlimit1}) and dropping higher order terms.  
Our criteria are satisfied by 
\begin{equation}\label{eq:GSRL2}
\Delta^{2}_{\cal R} = \Delta^{2}_{\cal R} |_{\rm GSRL}\left\{ [ 1 + {1 \over 4}I_1^2(k) + {1 \over 2}I_2(k)]^2 + {1 \over 2}I_1^2(k)\right\}
\end{equation}
where
\begin{eqnarray}\label{eqn:I1_I2}
I_1(k) & =  &{1 \over \sqrt{2} }\int_{0}^{\infty}{d\eta\over \eta}
G'(\ln \eta) X(k\eta)\,, \nonumber\\
I_2(k) & = &- 4 \int_{0}^{\infty} {d u \over u} [X + {1\over 3} X'] {f' \over f} F(u)\,,
\end{eqnarray}
with
\begin{equation}\label{eqn:F}
F(u) = \int_{u}^{\infty}{dv \over v^{2}} {f' \over f} \,.
\end{equation}
We call this the GSRL2 approximation.  
In the Appendix we discuss alternate
forms  \cite{Choe:2004zg}.

\begin{figure}[tbp]
\includegraphics[width=0.45\textwidth]{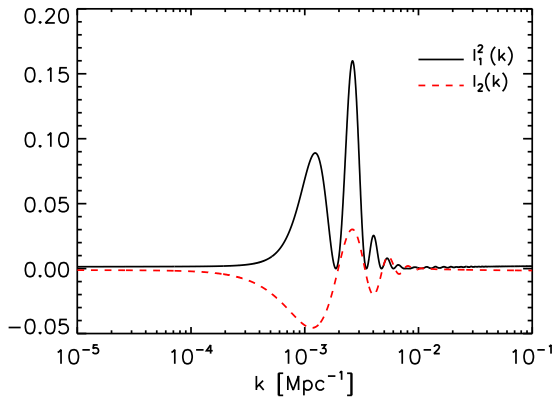}
\caption{Second order GSRL2 power spectrum correction functions $I_1^2$ and $I_2$ for the ML model.}
\label{plot:I1sq_I2_renormalized_ML}
\end{figure}

In the GSRL2 approach, $I_1$ corrections come half from the first order
calculation of the real part of the field and half from iterating the imaginary part to second order. 
In Fig. \ref{plot:I1sq_I2_renormalized_ML} we show $I_1^2$ and $I_2$ for the ML model.   Note that $I_1^2$ dominates
the correction to the net power as it always enhances power, while $I_2$ is both smaller
and oscillates in its correction.    Furthermore, both $|I_1^2| \ll 1$ and $|I_2| \ll 1$ for the ML
model which justifies a second order approach to these corrections.  The
GSRL2 correction can be taken to be $\{ 1 + I_1^2 + I_2 \}$ in this limit.

\begin{figure}[tbp]
\includegraphics[width=0.45\textwidth]{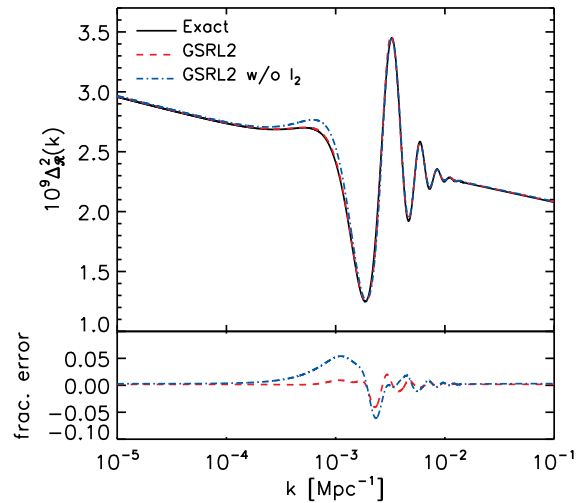}
\caption{GSRL2 approximation to the curvature power spectrum. Upper panel:  approximation of Eq.~(\ref{eq:GSRL2}) (red/dashed line) compared to the exact solution (black/solid line). We also show the GSRL2 approximation omitting the $I_2$ term (blue/dashed-dotted line). Lower panel:  fractional error between these GSRL2 approximations and  the exact solution.}
\label{plot:Powersp_Exact_GSRL2_GSRL2woI2}
\end{figure}

We show in Fig. \ref{plot:Powersp_Exact_GSRL2_GSRL2woI2} how the GSRL2 corrections reduce the  power spectrum errors of GSRL in Fig.~\ref{plot:GSRL_Muk_c0_bestfitmodel} for the ML model.   For the full GSRL2 expression
the power spectrum errors are reduced from the $10-20\%$ level to the $\lesssim 4\%$ level.
 We show that the GSRL2 approximation remains remarkably accurate for substantially
 larger features in the Appendix.

\begin{figure}[tbp]
\includegraphics[width=0.45\textwidth]{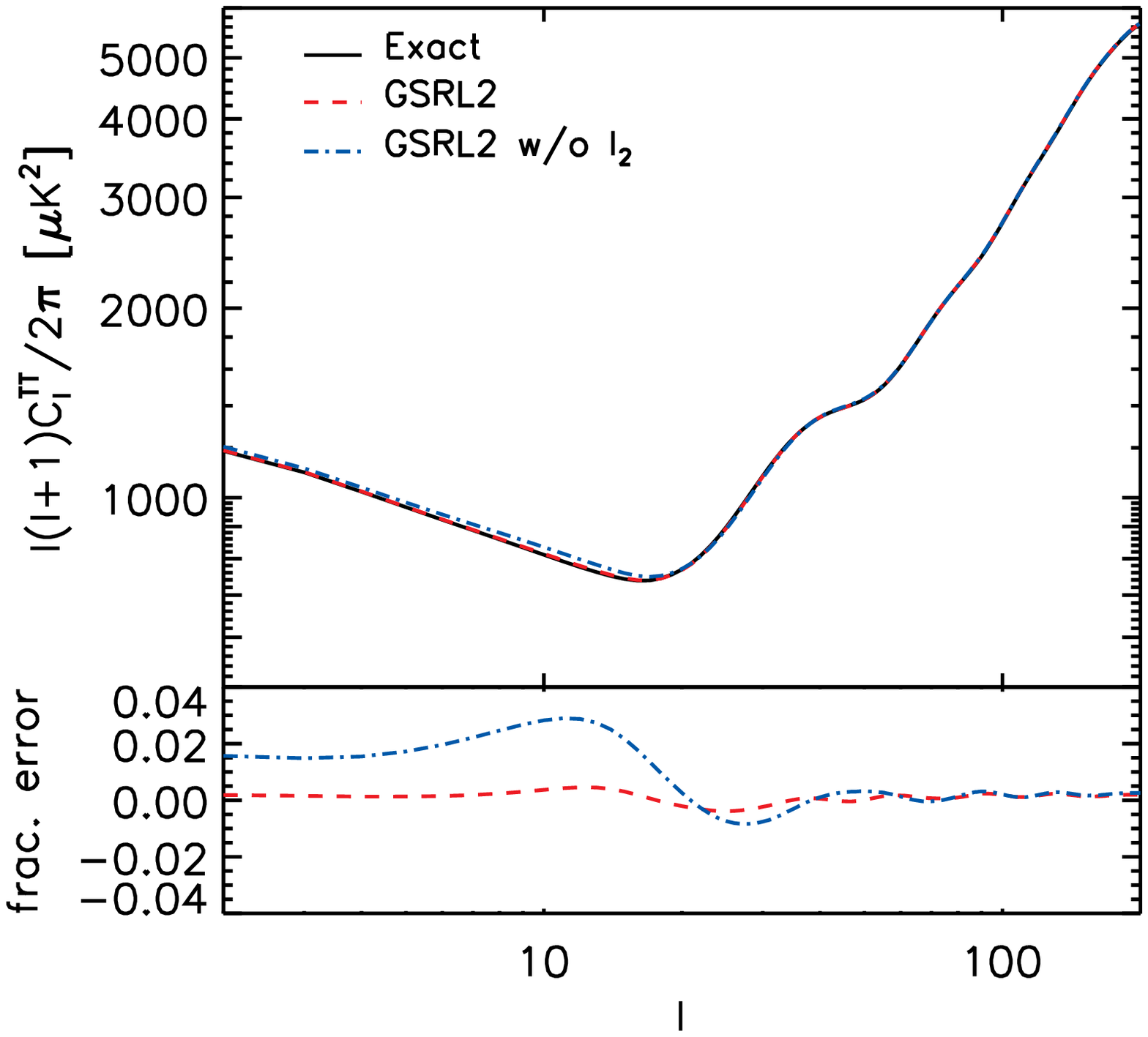}
\caption{GSRL2 approximation to the CMB temperature power spectrum. Upper panel:  approximation (red/dashed line) compared to the exact solution (black/solid line). We also show the GSRL2 approximation omitting the $I_2$ term (blue/dashed-dotted line). Lower panel:  fractional error between the GSRL2 approximations and the exact solution.}
\label{plot:ClTT_Exact_GSRL2_GSRL2woI2}
\end{figure}

\begin{figure}[tbp]
\includegraphics[width=0.45\textwidth]{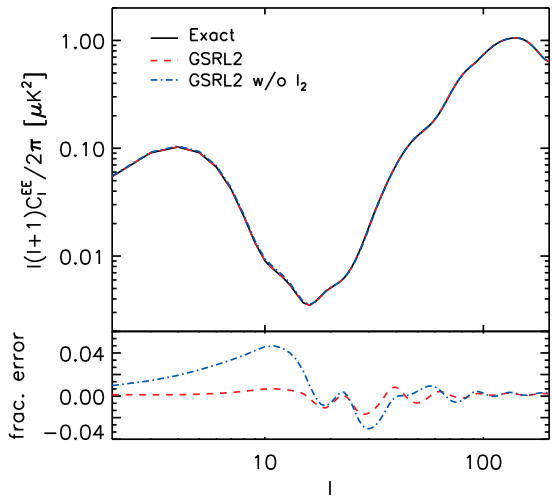}
\caption{GSRL2 approximation to the CMB $E$-mode polarization power spectrum.  The same as in Fig. \ref{plot:ClTT_Exact_GSRL2_GSRL2woI2}. }
\label{plot:ClEE_Exact_GSRL2_GSRL2woI2}
\end{figure}

Moreover, the errors are oscillatory and their observable consequence in the
CMB
is further reduced by projection.   The
temperature and polarization power spectra are shown in Fig.~\ref{plot:ClTT_Exact_GSRL2_GSRL2woI2}  and \ref{plot:ClEE_Exact_GSRL2_GSRL2woI2} 
and the errors are $\lesssim 0.5\%$ and $\lesssim 2\%$ for the respective spectra.

Given the intrinsic smallness of $I_{2}$ and its oscillatory nature,
 the most important correction comes
from the positive definite $I_1$ piece.   Note that it is a single integral over  the same $G'$
function as in the linear case.  
Thus, $I_{1}$ corrections simply generalize the GSRL
mapping between $G'$ and curvature
in a manner that is equally simple to calculate.
  $I_2$ on the other hand is more complicated and involves a non-trivial
double integral with a different dependence on the inflaton potential. 

 We also show in 
 Figs.~\ref{plot:Powersp_Exact_GSRL2_GSRL2woI2}-\ref{plot:ClEE_Exact_GSRL2_GSRL2woI2} 
the results for the GSRL2 expression
with $I_2$ omitted.  While the curvature power spectrum errors increase slightly
to $\sim 5\%$, the temperature power spectrum errors at   $\lesssim 2\%$  are  still well below the $\sim 20\%$
cosmic variance errors per $\ell$ at $\ell \sim 30$.  They  are
even sufficient for the cosmic variance limit of coherent
deviations across the full range of the feature ($20 \lesssim \ell \lesssim 40$)  $20\%/\sqrt{20} \sim 4-5\%$ in the ML case.  The polarization spectrum has slightly larger
errors due to the reduction of projection effects but still satisfies these cosmic
variance based criteria.

\section{Applications}
\label{sec:applications}

In the previous section, we have shown that a particular variant of the GSR approximation
which we call GSRL2 provides a non-linear mapping between $G'$ and the curvature power spectrum. $G'$ quantifies
the deviations from slow roll in the background and moreover is 
to good approximation directly related to the inflaton potential.  These relations
remain true even
when the slow-roll parameter $\eta_H$ is not small compared to unity for a fraction of
an $e$-fold.  

This relationship is useful for considering inflation-model independent constraints
on the inflaton potential.  It is likewise useful for inverse or model building
approaches of finding inflaton potential classes that might fit some observed
feature in the data.   We intend to further explore these applications in a future work.

\begin{figure}[tbp]
\includegraphics[width=0.45\textwidth]{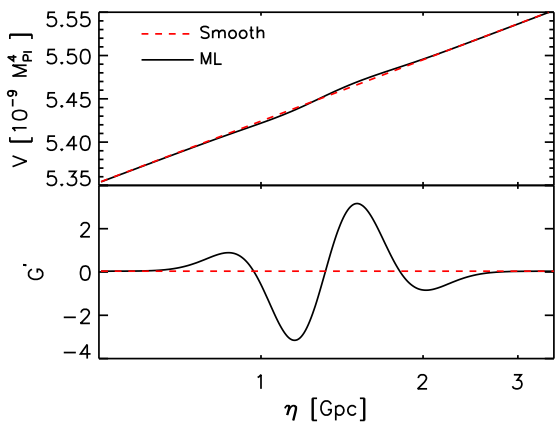}
\caption{Alternate inflationary model with a perturbation in the mass.  Upper panel: 
comparison of potential in Eq.~(\ref{eqn:alternative_V})  (black/solid line) and the smooth potential (red/dashed line). Lower panel: source function of the deviation from slow roll $G^{\prime}$ for the same models.}
\label{plot:Gprime_V_exp}
\end{figure}

Here as a simple example let us consider a potential that differs qualitatively from the
step potential but shares similar observable properties through $G'$: $V(\phi)=m^2_{\rm eff}\phi^2/2$ where the effective mass of the inflaton now has a transient perturbation instead
of a step
\be\label{eqn:alternative_V}
m^2_{\rm eff} = m^2\left[1+Ae^{-(\phi-b)^2/(2\sigma^2)}(\phi-b)\right]
\ee
In Fig. \ref{plot:Gprime_V_exp} we show the potential for the choice of parameters $b=14.655$, $A=0.0285$, $\sigma=0.025$, and $m=7.126\times10^{-6}$ (upper panel) and we also show $G^{\prime}$ in the lower panel. For comparison we show the smooth case $A=0$.  Comparison with Figs.~\ref{plot:potential_time_bestfitmodel} and 
\ref{plot:slow_roll_parameters_bestfitmodel} shows that this potential, which has
a bump and a dip instead of a step, produces a similar main feature in $G'$ but
has additional lower amplitude secondary features. 

\begin{figure}[tbp]
\includegraphics[width=0.45\textwidth]{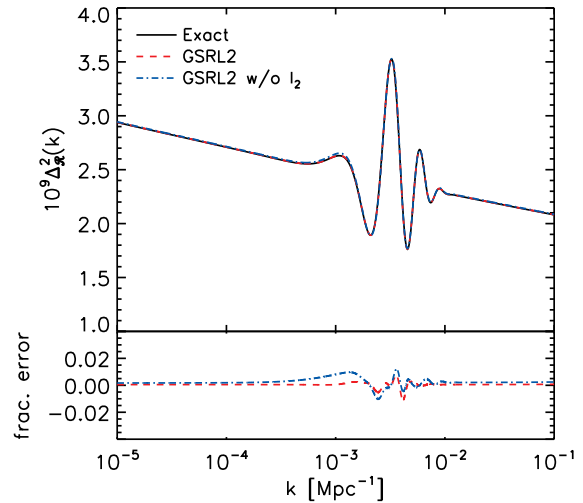}
\caption{GSRL2 approximation to the alternate model of Fig.~\ref{plot:Gprime_V_exp}.
Upper panel: approximation (red/dashed line) compared to the exact solution (black/solid line) for an effective mass given by Eq.~(\ref{eqn:alternative_V}). We also show the GSRL2 approximation with $I_2$ omitted (blue/dashed-dotted line). Lower panel: fractional error between GSRL2 approximations and the exact solution.}
\label{plot:Pofk_GSRL2_Muk_exp_similarGprime} 
\end{figure}

In Fig. \ref{plot:Pofk_GSRL2_Muk_exp_similarGprime} we compare the GSRL2 approximation with and without the double integral $I_{2}$ term 
compared to the exact solution.  Notice that GSRL2 performs equally well for this very different sharp potential
feature.   Furthermore, similarity in $G'$ with the step potential carries over to similarity
in the curvature power spectrum.

\section{Discussion}
\label{sec:discussion}

We have shown that a variant of the generalized slow roll (GSR) approach remains 
percent level accurate at predicting  order unity deviations in the
observable CMB temperature and polarization power spectra from sharp potential
features.   Unlike other variants which explicitly require
 $|\eta_{H}|\ll 1$, and hence nearly constant $\epsilon_{H}$, our approach allows
 $\eta_{H}$ to be order unity, as long as it remains so for less than an $e$-fold, and hence $\epsilon_{H}$ to
 vary significantly.  We have tested our GSR variant against a step function model
 that has been proposed to explain features in the CMB temperature power spectrum
 at $\ell \sim 20-40$.

Our analysis also shows that to good approximation
 a single function, $G'(\ln \eta)$, controls the
observable features in the curvature power spectrum even in the presence of large
features.  
We have explicitly checked this relationship and the robustness of our approximation 
by constructing two different inflationary
models with similar $G'$.

Therefore observational constraints from the CMB can be mapped 
directly to constraints on this function independently of the model for inflation.  
Moreover, this function is also simply related to the slope and curvature of the
inflaton potential in the same way that scalar tilt is related to the potential
in ordinary slow roll $G' \approx 3 (V_{,\phi}/V)^{2} - 2 (V_{,\phi\phi}/V)$.
These model independent constraints can then be simply interpreted in
terms of the inflation potential.   We intend to explore these applications in a future
work.

{\it Acknowledgments:}  We thank Hiranya Peiris for sharing code used to
crosscheck exact results.   We thank Michael Mortonson, Kendrick Smith and Bruce Winstein for useful conversations.
This work was supported by the KICP under
NSF contract PHY-0114422.   WH was additionally supported by DOE contract DE-FG02-90ER-40560 and the Packard Foundation.

\appendix

\begin{figure}[tbp]
\includegraphics[width=0.45\textwidth]{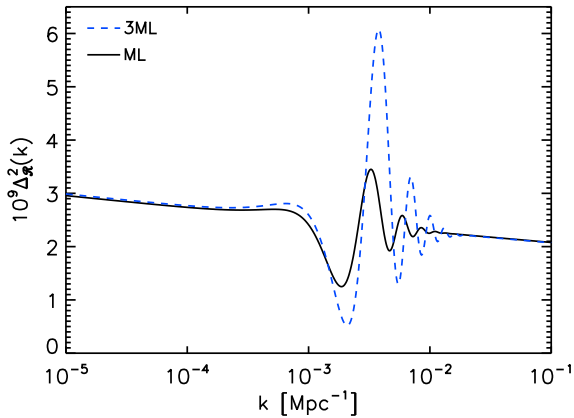}
\caption{Curvature power spectrum for the ML and 3ML models.}
\label{plot:ExactPowersp_ML_3ML}
\end{figure}

\section{Other GSR Variants}

In this Appendix, we compare various alternate forms  discussed in the literature for the curvature power spectrum
under the GSR approximation.    We test these
approximations against the GSRL and GSRL2 approximations of the main text
for the ML model and a more extreme case with $c= 3 c_{\rm ML}=0.004515$ (with other
parameters fixed)
denoted 3ML (see Fig.~\ref{plot:ExactPowersp_ML_3ML}).  We begin by considering variants that are linear in
the GSR approximation and then proceed to second order iterative approaches.

\begin{figure}[tbp]
\includegraphics[width=0.45\textwidth]{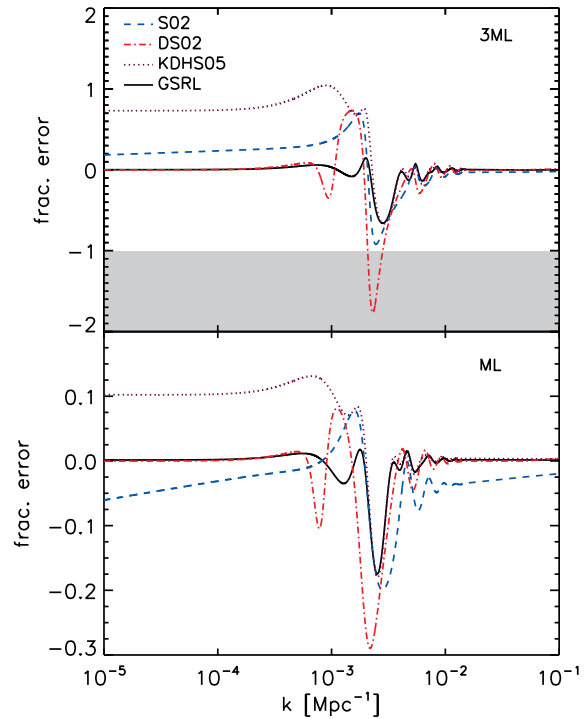}
 \caption{Fractional error in the curvature power spectrum for first order GSR variants
 for the ML model (lower) and the 3ML model (upper).}
 \label{plot:linearized_variants}
\end{figure}

The first variant is the original linearized form of GSRS given in 
\cite{Stewart:2001cd} (``S02")
\be\label{eqn:Stewart74}
\Delta^2_{\cal R}(k) = {1\over f^2}\left[1+{2\over 3}{f^{\prime}\over f} + {2\over 3}\int_x^{\infty}{du\over u}W(u)g(\ln u)\right] \,.
\ee
Like GSRS, this approximation depends on an arbitrary choice
of $x$ but its impact is exacerbated by the linearization of the correction here.   In Fig.~\ref{plot:linearized_variants} we show the fractional error in this approximation for
$\eta_{\rm min}=10^{-1}$ Mpc.  Note that because of the linearization, the curvature power
spectrum can reach the unphysical negative regime (shaded region).

A second variant further exploits the relationship between the GSR source
functions $f$, $f'/f$ and $g$ and the potential through the slow-roll parameters 
(see Eq.~(\ref{eqn:fslowroll})). 
By further assuming that  $|\eta_H|  \ll 1$, terms involving $V_{,\phi}/V$ can be taken to
be constant and evaluated instead at horizon crossing $k=aH$ (see Eq.~(\ref{eqn:epsevol})). 
Finally by rewriting the change in $f'/f$ as the integral of $(f'/f)'$, one obtains
\cite{Dodelson:2001sh} (``DS02")
\ba
\Delta^2_{\cal R}(k) &=& {V \over 12\pi^2} \left( {V\over V_{,\phi} }\right)^2\Big\{1+(3\alpha-{1\over6})({V_{,\phi}\over V})^2|_{k=aH}\nn \\
&& -2\int_0^{\infty}{du\over u}W_\theta(1,u){V_{,\phi\phi}\over V}\Big\},
\label{eq:SteGSR}
\ea
where $\alpha \approx 0.73$
and with $\eta \approx 1/aH$, $u = k/aH$.
Here
\begin{equation}
W_\theta(u_*,u) = W(u)-\theta(u_*-u) 
\label{eqn:wtheta}
\end{equation}
with the step function  $\theta(x)=0$ for $x<0$ and $\theta(x)=1$ for $x\ge0$.
Note that $\lim_{u\rightarrow 0} W_\theta(1,u) =0$ and hence the function has weight only near horizon
crossing at $u \approx 1$.

For cases like the ML and 3ML models where $\eta_H$ is neither small nor smoothly
varying, these DS02 
assumptions have both positive and negative consequences.  
They largely solve the problem for superhorizon modes discussed in \S \ref{sec:GSRL} by extrapolating the evaluation of the potential terms from $k\eta \ll 1$ to $k\eta \sim 1$.  
On the other hand, a large $\eta_H$ means that $\epsilon_H$ evolves significantly.
Artifacts of this evolution appear through the prefactor $(V/V_{,\phi})^2 \propto 1/\epsilon_{H}$  in 
Eq.~(\ref{eq:SteGSR}) most notably in the form of a spurious feature at $k\sim 10^{-3}$ 
Mpc$^{-1}$ in Fig.~\ref{plot:linearized_variants}.  Finally, like S02, DS02 does
not guarantee a positive definite power spectrum.

A third variant is to replace $G'$ with $2g_V/3$ in Eq. (\ref{eqn:GSR2}) so that
the source directly reflects the potential  \cite{Kadota:2005hv}
(``KDHS05")
\be
\ln\Delta_{\cal R}^2(k) = G(\ln\eta_{\rm min}) +{2\over 3} \int_{\eta_{\rm min}}^{\infty}{d\eta\over\eta}W(k\eta)g_V \,.
\label{eqn:GSRV}
\ee
As we have seen in \S \ref{sec:GSRL}, this approximation is actually fairly good {\it locally}
in $\ln\eta$ and hence locally in $k$ around the feature.
 However the omission of $\eta_H^2$ terms causes a net error in the spectrum for
 $k$ modes that cross out of the horizon before the inflaton reaches the feature. Hence
like the GSRS approximation, KDHS05  overpredicts power at low $k$ for the ML and 3ML models.
 Fig.~\ref{plot:linearized_variants} shows a choice with $\eta_{\rm min}=10^{-1}$ Mpc.

\begin{figure}[tbp]
\includegraphics[width=0.45\textwidth]{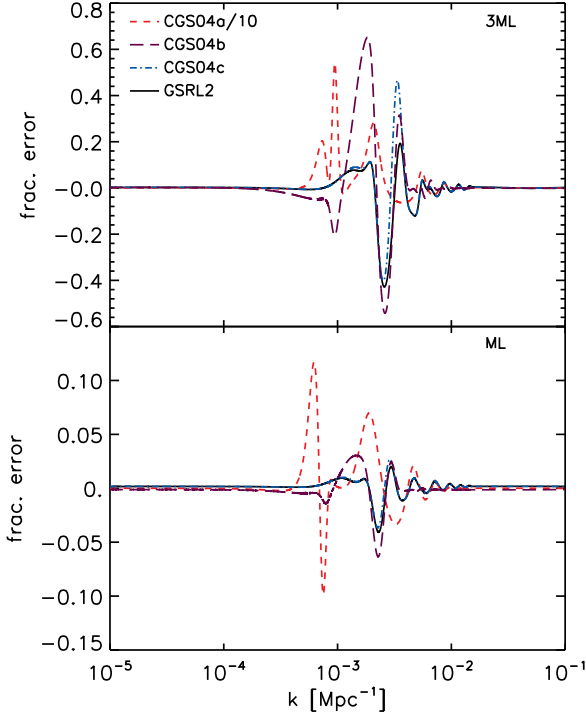}
\caption{Fractional error in the curvature power spectrum for second order GSR variants
(see text)
for the ML model (lower panel) and the 3ML model (upper panel). Note that
the error in CGS04a has been divided by a factor of 10 for plotting purposes.}
\label{plot:second_order_variants_combined_v2}
\end{figure}

\begin{figure}[tbp]
\includegraphics[width=0.45\textwidth]{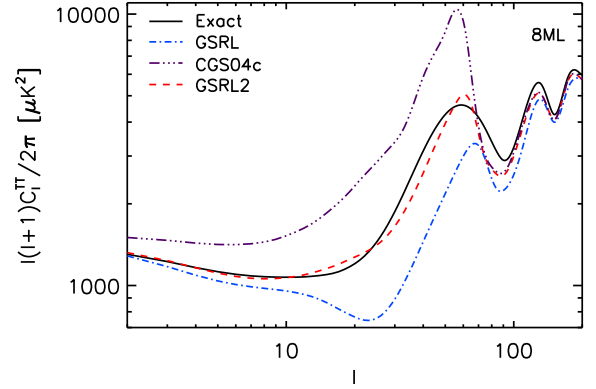}
\caption{Temperature power spectrum for $c=8 c_{\rm ML}$ (and the other parameters fixed).
Note that even in this extreme, observationally unviable, case
the temperature power spectrum has $< 22\%$ errors everywhere for GSRL2 whereas
the linear GSRL and CGS04c approximations substantially
under and over predict power respectively.}
\label{plot:ClTT_8ML_GSRL_GSRL2_CGS04c}
\end{figure}

We consider next second order GSR variants.  
The first variant \cite{Choe:2004zg} begins with a second order approach as in GSRL2
but then further assumes that functions such as $f'/f$ can be
approximated by a Taylor expansion around $x_*$ to obtain
(``CGS04a")
\ba
\Delta^{2}_{\cal R} &=& {1\over f_*^2}\Big\{1-2\alpha_*{f'_*\over f_*} + \left(-\alpha_*^2+{\pi^2\over12}\right){f^{''}_*\over f_*} \nonumber\\
&& + \left(3\alpha_*^2-4+{5\pi^2\over 12}\right)\left({f'_*\over f_*}\right)^2 \nonumber\\
&&+ \left[-{1\over 3}\alpha_*^3+{\pi^2\over 12}\alpha_* - {4\over 3}+{2\over 3}\zeta(3)\right]{f^{\prime\prime\prime}_* \over f_*} \nonumber\\
&& + \left[3\alpha_*^3-8\alpha_*+{7\over 12}\pi^2\alpha_*+4-2\zeta(3)\right]{f'_*f^{''}_*\over f^2_*} \nonumber\\
&& + A\left({f'_*\over f_*}\right)^3\Big\},
\ea
where $A=-4\alpha_*^3+16\alpha_*-5/3\pi^2\alpha_*-8+6\zeta(3)$, $\zeta$ is the Riemann zeta function, and $\alpha_*=\alpha-\ln(x_*)$.  This approach is essentially a standard slow
roll approximation carried through to third order with the help of an exact solution for
power law inflation.   For the ML and 3ML models, applying this approximation leads
to qualitatively incorrect results as one might expect. We show this variant in Fig.~\ref{plot:second_order_variants_combined_v2} with $x_*=1$.

A second variant attempts to retain both the generality of GSR and the evaluation of
central terms at horizon crossing by implicitly modifying terms of order $(f'/f)^{3}$ and
higher when compared with GSRL2 \cite{Choe:2004zg} (``CGS04b")
\ba
\ln\Delta^{2}_{\cal R} &=& \ln\left({1\over f_*^2}\right) + {2\over 3}{f'_*\over f_*} + {1\over 9}\left({f'_*\over f_*}\right)^2 \\
&& + {2\over 3} \int_0^\infty{du\over u}W_\theta(u_*,u)g(u)  \nonumber\\
&& + {2\over 9}\left[\int_0^\infty{du \over u} X(u)g(u)\right]^2 \nonumber\\
&& - {2\over 3}\int_0^\infty {du\over u} X(u)g(u)\int_u^\infty {dv\over v^2}g(v)\nonumber\\
&& - {2\over 3}\int_0^\infty {du\over u} X_\theta(u_*,u)g(u)\int_u^\infty {dv\over v^4}g(v)\,, \nonumber
\ea
where $W_\theta$ was given in Eq.~(\ref{eqn:wtheta}) and 
\ba
X_\theta(u_*,u) = X(u)-{u^3\over 3}\theta(u_*-u)\,.
\ea
Here, the subscript $*$ denotes evaluation near horizon crossing. In Fig.~\ref{plot:second_order_variants_combined_v2}
we show the result with $u_*=1$.      Notably it performs worse than the first order GSRL approximation
for the  3ML model.

Finally, the last variant considered takes \cite{Choe:2004zg} (``CGS04c")
\ba
\label{eqn:cgs04c}
\ln\Delta^{2}_{\cal R} &=& -\int_0^\infty {du\over u} W'(u)\left[\ln\left({1\over f^2}\right)+{2\over 3}{f'\over f}\right]\\
&& + 2\left[\int_0^\infty{du \over u}\left(X(u)+{1\over 3}X'(u)\right){f'\over f}\right]^2 \nonumber\\
&& -4\int_0^\infty{du\over u}\left(X(u)+{1\over 3}X'(u)\right){f'\over f}F(u), \nonumber
\ea
where $F(u)$ is given by Eq. (\ref{eqn:F}).
CGS04c is closely related to GSRL2 as integration by parts shows
\begin{equation}
\Delta^{2}_{\cal R} = \Delta^{2}_{\cal R} |_{\rm GSRL}e^{I_1^2(k) + I_2(k)} \,.
\end{equation}
The main difference is that the second order corrections are exponentiated.   This
causes a noticeable overcorrection for the 3ML model when compared with GSRL2.
In Fig. \ref{plot:second_order_variants_combined_v2} we compare the three variants mentioned above.

   Furthermore, in spite
of the $20-40\%$ errors in the curvature power spectrum in the 3ML model for
GSRL2, the CMB temperature power spectrum has only $1-2\%$ errors for $\ell \ge 20$ 
and a maximum of $< 5\%$ errors at $\ell<20$.  As discussed in the text, this level
of error is sufficient for even cosmic variance limited measurements at the $\ell \lesssim 40$
multipoles of the feature.  
This reduction
is due to the oscillatory nature of the curvature errors and projection effects in 
temperature. 

In fact for even larger deviations GSRL2 still performs surprisingly well for
the temperature power spectrum. In Fig.~\ref{plot:ClTT_8ML_GSRL_GSRL2_CGS04c} we show the temperature power spectra for the GSRL2 approximation, and compare it with GSRL and CGS04c for a very extreme case with $c=8 c_{\rm ML}$ (and the other parameters fixed). GSRL2 has a maximum of $22\%$ error in the temperature power spectrum and predicts qualitatively correct features.    Finally,  the dominant correction is
from the term that is quadratic in $I_1$.  The simplified GSRL2 form of 
\begin{equation}
\Delta^{2}_{\cal R} = \Delta^{2}_{\cal R} |_{\rm GSRL}[1+{I_1^2(k)}] \,,
\end{equation}
works nearly as well.   Thus, the curvature power spectrum still depends only on $G'$
to good approximation even in the most extreme case.

\vfill

\bibliographystyle{arxiv_physrev}

\bibliography{GSR}

\def\eprinttmppp@#1arXiv:@{#1}
\providecommand{\arxivlink[1]}{\href{http://arxiv.org/abs/#1}{arXiv:#1}}
\providecommand{\arxivlinknopre[1]}{\href{http://arxiv.org/abs/#1}{#1}}
\providecommand{\eprintmod}[1][XXXX.XXXX]{\IfSubStr{#1}{arXiv}{\arxivlinknopre%
{#1}}{\arxivlink{#1}}}
\providecommand{\adsurl}[1]{\href{#1}{ADS}}
\begin{thebibliography}{18}
\expandafter\ifx\csname natexlab\endcsname\relax\def\natexlab#1{#1}\fi
\expandafter\ifx\csname bibnamefont\endcsname\relax
  \def\bibnamefont#1{#1}\fi
\expandafter\ifx\csname bibfnamefont\endcsname\relax
  \def\bibfnamefont#1{#1}\fi
\expandafter\ifx\csname citenamefont\endcsname\relax
  \def\citenamefont#1{#1}\fi
\expandafter\ifx\csname url\endcsname\relax
  \def\url#1{\texttt{#1}}\fi
\expandafter\ifx\csname urlprefix\endcsname\relax\def\urlprefix{URL }\fi

\bibitem{Lidsey:1995np}
J.~E. Lidsey {\em et~al.},
\newblock Rev. Mod. Phys. {\bf 69}, 373 (1997), [\eprintmod[astro-ph/9508078]].

\bibitem{Bennett:2003bz}
WMAP, C.~L. Bennett {\em et~al.},
\newblock Astrophys. J. Suppl. {\bf 148}, 1 (2003),
  [\eprintmod[astro-ph/0302207]].

\bibitem{Peiris:2003ff}
WMAP, H.~V. Peiris {\em et~al.},
\newblock Astrophys. J. Suppl. {\bf 148}, 213 (2003),
  [\eprintmod[astro-ph/0302225]].

\bibitem{Covi:2006ci}
L.~Covi, J.~Hamann, A.~Melchiorri, A.~Slosar and I.~Sorbera,
\newblock Phys. Rev. {\bf D74}, 083509 (2006), [\eprintmod[astro-ph/0606452]].

\bibitem{Hamann:2007pa}
J.~Hamann, L.~Covi, A.~Melchiorri and A.~Slosar,
\newblock Phys. Rev. {\bf D76}, 023503 (2007), [\eprintmod[astro-ph/0701380]].

\bibitem{Mortonson:2009qv}
M.~J. Mortonson, C.~Dvorkin, H.~V. Peiris and W.~Hu,
\newblock Phys. Rev. {\bf D79}, 103519 (2009), [\eprintmod[0903.4920]].

\bibitem{Pahud:2008ae}
C.~Pahud, M.~Kamionkowski and A.~R. Liddle,
\newblock Phys. Rev. {\bf D79}, 083503 (2009), [\eprintmod[0807.0322]].

\bibitem{Joy:2008qd}
M.~Joy, A.~Shafieloo, V.~Sahni and A.~A. Starobinsky,
\newblock JCAP {\bf 0906}, 028 (2009), [\eprintmod[0807.3334]].

\bibitem{Adams:2001vc}
J.~A. Adams, B.~Cresswell and R.~Easther,
\newblock Phys. Rev. {\bf D64}, 123514 (2001), [\eprintmod[astro-ph/0102236]].

\bibitem{Hunt:2004vt}
P.~Hunt and S.~Sarkar,
\newblock Phys. Rev. {\bf D70}, 103518 (2004), [\eprintmod[astro-ph/0408138]].

\bibitem{Habib:2004kc}
S.~Habib, A.~Heinen, K.~Heitmann, G.~Jungman and C.~Molina-Paris,
\newblock Phys. Rev. {\bf D70}, 083507 (2004), [\eprintmod[astro-ph/0406134]].

\bibitem{Joy:2005ep}
M.~Joy, E.~D. Stewart, J.-O. Gong and H.-C. Lee,
\newblock JCAP {\bf 0504}, 012 (2005), [\eprintmod[astro-ph/0501659]].

\bibitem{Kadota:2005hv}
K.~Kadota, S.~Dodelson, W.~Hu and E.~D. Stewart,
\newblock Phys. Rev. {\bf D72}, 023510 (2005), [\eprintmod[astro-ph/0505158]].

\bibitem{Stewart:2001cd}
E.~D. Stewart,
\newblock Phys. Rev. {\bf D65}, 103508 (2002), [\eprintmod[astro-ph/0110322]].

\bibitem{Choe:2004zg}
J.~Choe, J.-O. Gong and E.~D. Stewart,
\newblock JCAP {\bf 0407}, 012 (2004), [\eprintmod[hep-ph/0405155]].

\bibitem{Dodelson:2001sh}
S.~Dodelson and E.~Stewart,
\newblock Phys. Rev. {\bf D65}, 101301 (2002), [\eprintmod[astro-ph/0109354]].

\bibitem{Gong:2005jr}
J.-O. Gong,
\newblock JCAP {\bf 0507}, 015 (2005), [\eprintmod[astro-ph/0504383]].

\end{thebibliography}

\end{document}